# EVANESCENT WAVES AND SUPERLUMINAL BEHAVIOR OF MATTER

LUCA NANNI

Faculty of Natural Science, University of Ferrara, Saragat 1 – 44122 Ferrara, Italy

luca.nanni@student.unife.it

**Abstract** An evanescent wave is a non-propagating wave with an imaginary wave vector. In this study, we prove that these are solutions of the tachyon-like Klein–Gordon equation, and that in the tunneling of ultrarelativistic spin-1/2 particles they describe superluminal states arising from interactions between a particle and barrier. These states decay as a particle emerges from the opposite side of a potential barrier, conserving the same initial energy but not necessarily the same mass. The obtained theory is applied to the neutrino, to explain flavor oscillations during free flight and determine the conditions that maximize the probability of their occurrence.

*Keywords*: Evanescent waves; quantum tunneling; relativistic wave packet; neutrino; mass oscillation.

## 1. Introduction

Evanescent waves are localized waves that have the property of resisting diffraction in dispersive media, such as potential barriers, even over long distances [1-8]. These non-propagating waves are characterized by imaginary wave vectors [1], and decay exponentially along the direction of propagation within the barrier. In optics and quantum mechanics, evanescent waves are associated with tunneling phenomena, where a wave packet collides with a potential barrier, and emerges from the opposite side with reduced amplitude but the same energy as the incoming wave packet [9-12], as illustrated in Fig. 1.

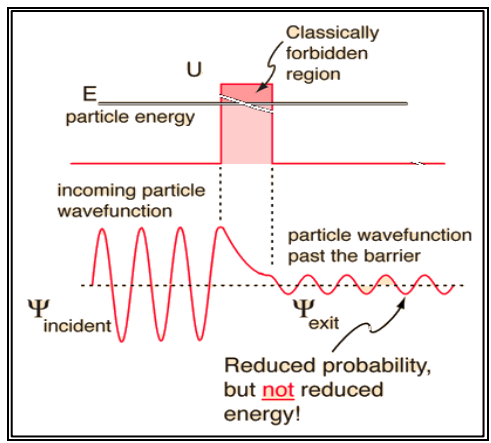

**Figure 1**: The incident wave meets the potential barrier, and its amplitude decreases asymptotically. The outgoing wave emerges with reduced amplitude but the same energy.

Within a barrier, a photon or the massive particle behaves like an evanescent wave, which is transmitted with superluminal group velocity if the barrier is sufficiently large [13-16]. This phenomenon is known as the Hartman effect [17] and is based on the fact that the tunneling time, computed as the Wigner phase time, does not depend on the barrier dimension [18-20]. This effect has been experimentally verified in optics [21] and cosmology [22]. On the other hand, in particle physics the Hartman effect has been extensively investigated but has not yet been supported by experimental evidence [23-25]. However, experimental confirmation could come from neutrino physics (specifically superluminality and flavor oscillations) [26].

In this work, we will prove that spinor evanescent waves are solutions of the tachyon-like Dirac equation (Lemke equation) [27], and that their components are solutions of the associated tachyon-like Klein–Gordon (KG) equation. This means that evanescent waves in spinorial form are adequate to describe spin-1/2 particles in the tachyonic regime. Then, we will investigate the behavior of an ultrarelativistic spin-1/2 fermion in a potential barrier, highlighting that tunneling can promote the particle into an *excited* state characterized by an imaginary mass and tachyonic energy. Upon leaving the barrier, the particle returns to the ultrarelativistic regime, returning the energy taken during the interaction with the barrier. During this process, there is a non-zero probability that the fermion can change mass its and momentum, while always conserving the energy of the initial particle.

The obtained results will be applied to a high-energy neutrino, to explain its flavor oscillation during free flight. More precisely, we will determine the particle energy and barrier dimensions for which the probability that the neutrino changes its flavor is maximized. This theory is based on the fact that a neutrino interacts with surrounding matter so as to create a dispersive medium that acts as a potential barrier. If these interactions are sufficiently intense, then the neutrino *becomes* superluminal, making it possible to change its flavor. The problem is that the neutrino interacts very weakly with matter, and therefore the expected probability that it is in a superluminal *excited* state is very low, but still different from zero. Therefore, this study suggests that the superluminal behavior of the neutrino may be studied through investigating its flavor oscillations during free flight.

## 2. Evanescent Waves and the Tachyon-like Dirac Equation

Within the potential barrier $U > E$, where E is the energy of the initial particle, the De Broglie wave vector is given by [8]

$$\chi = \frac{\sqrt{2m(E-U)}}{\hbar} = \frac{i\sqrt{2m(U-E)}}{\hbar}. \tag{1}$$

Because such waves are associated with superluminal motions in tunneling phenomena, we must prove that Eq. (1) is the wave vector of a tachyon, which is a particle with imaginary mass. To this end, we set the term $(U - E)$ equal to the tachyonic energy:

$$(U - E) = \gamma_t mc^2, \tag{2}$$

where $\gamma_t$ is the tachyonic relativistic factor [28]:

$$\gamma_t = \frac{1}{\sqrt{\left(\frac{u_t}{c}\right)^2 - 1}} \quad where \ u_t > c. \tag{3}$$

Substituting Eq. (2) into (1), we obtain

$$\chi = \frac{i\sqrt{2m(\gamma_t mc^2)}}{\hbar} = \frac{\sqrt{2\gamma_t (im)^2 c^2}}{\hbar}, \tag{4}$$

which is simply the De Broglie wave vector of a tachyonic plane wave related to a particle with imaginary mass [21, 29].

Let us suppose that the free particle arriving at the barrier is a monochromatic Dirac spinor. For further simplification, we also suppose that the particle propagates along the x-axis of the reference frame. In this manner, there is no spin-flip, and the spinor has only two components [30]. Each component of the Dirac spinor must separately satisfy the KG equation:

$$|\psi\rangle = (\psi_1, \psi_2) \quad \Rightarrow \quad \partial_x^2 \psi_i - \partial_t^2 \psi_i = \left(\frac{mc}{\hbar}\right)^2 \psi_i. \tag{5}$$

It is known that for a free particle, solutions of the KG equation are plane waves:

$$\varphi(x,t) = exp\{i(kx - \omega t)\}, \tag{6}$$

where $\omega$ is given by

$$\omega = c\sqrt{k^2 + (mc/\hbar)^2}. \tag{7}$$

If $|\omega| < (mc^2/\hbar)$, then the wave is evanescent in space [1]. We may use Eq. (7) to calculate $\omega$ in the case where the wave vector $k$ is that of a tachyonic particle:

$$\omega_t = c\sqrt{\chi^2 + (imc/\hbar)^2} = c\sqrt{(1 + 2\gamma_t)(imc/\hbar)^2}. \tag{8}$$

For a spin-1/2 free particle propagating along the x-axis, the Dirac spinor is [31]

$$|\psi\rangle = \begin{pmatrix} 1 \\ \frac{\gamma}{\gamma+1} \end{pmatrix} exp\{i(kx - \omega t)\}. \tag{9}$$

where $\gamma$ is the Lorentz factor. Once the particle, is within the potential barrier in probabilistic terms, it is described by the following evanescent spinor:

$$|\varphi_{ev.}\rangle = \begin{pmatrix} 1 \\ 1 \end{pmatrix} exp\left\{\frac{mc}{\hbar}(1 + 2\gamma_t)^{1/2}(ct - x)\right\}, \tag{10}$$

obtained by replacing $\chi$ and $\omega_t$ in Eq. (9). Because the evanescent wave is localized and does not propagate in time, the second component of the spinor becomes unitary, and the temporal term $ct$ may be omitted. Moreover, considering that the particle exits the barrier as a Dirac wave with a dumped amplitude, it is reasonable to expect that the evanescent wave inside the barrier is compatible with a (virtual) tachyon with a speed $u_t$ slightly higher than that of light.

We must now prove that the components of the spinor (10) are solutions of the tachyon-like KG equation, compatible with the momentum relation

$$E^2 = p^2 c^2 - m^2 c^4$$

The tachyon-like KG equation is given by [27]

$$\partial_x^2 \psi_i - \partial_t^2 \psi_i = -\left(\frac{mc}{\hbar}\right)^2 \psi_i. \tag{11}$$

Substituting the explicit form of each component of the evanescent wave (9) into Eq. (11), eventually also including the time part, we can easily verify the equality between the right- and left-hand sides. Thus, we have shown that the evanescent wave describes the superluminal behavior of matter in tunneling phenomena.

On the basis of the obtained results, we can state that quantum tunneling *brings* the particle into a tachyonic state, which we define from now on as a *localized excited state* (virtual). This result is completely consistent with the Hartman effect, where the particle superluminality is estimated through the time required to cross the potential barrier. This process occurs thanks to the interaction between the particle and barrier: the higher the barrier, the greater the probability that the particle passes into a tachyonic state. This conclusion is consistent with the fact that the Hartman effect is particularly favored by sufficiently large barriers [17, 23]. In fact, in this situation the probability of interaction between the particle and barrier increases, which is precisely the condition necessary to obtain a superluminal state.

**3. Tunneling of Ultrarelativistic Spin-1/2 Particles**

In the previous section, we considered for simplicity a monochromatic ultrarelativistic Dirac wave arriving at a potential barrier. However, according to the uncertainty principle a free particle propagating in space is represented by a wave packet that progressively spreads [24]. Therefore, to continue this study we reconsider the tunneling phenomenon of an ultrarelativistic Dirac particle starting from a wave packet arriving at a potential barrier with $U$ being slightly higher than the initial particle energy. For this purpose, we can choose an ultrarelativistic Gaussian wave packet, given a plane wave modulated by a Gaussian function [32]:

$$\psi_G(x,t) = \begin{pmatrix} 1 \\ \frac{\gamma}{\gamma+1} \end{pmatrix} exp\{-i(kx - \omega t)\} \frac{exp\{-(x-ut)^2/4\sigma_x^2[(1+it)/2\gamma m\sigma_x^2]\}}{(2\pi)^{3/4}\sqrt{2m}\sigma_x^3[(1+it)/2\gamma m\sigma_x^2]^{3/2}}, \qquad (12)$$

where $u$ is the initial particle velocity, $\gamma$ is the Lorentz factor, and $\sigma_x$ is the spread indicator. As the Gaussian packet meets the barrier, each plane wave that forms the envelope is reshaped in an evanescent wave [1]. Overall, within the barrier we have a monochromatic evanescent wave, which can be represented as

$$\varphi_G(x,t) = \begin{pmatrix} 1 \\ 1 \end{pmatrix} e^{-\chi x} e^{-\omega t}. \qquad (13)$$

As usual, the temporal part of the evanescent wave may be omitted. If we denote by $x_0$ the point where the barrier of length $L$ begins, then in the space with $x < x_0$ the total spinor is given by a linear combination of the incident wave and the part of the wave that is reflected and does not penetrate the barrier [23]:

$$\psi_{x<x_0}(x,t) = \begin{pmatrix} 1 \\ \frac{\gamma}{\gamma+1} \end{pmatrix} \Phi(u,\sigma_x)[exp\{-i(kx-\omega t)\} + c_R exp\{i(kx-\omega t)\}], \qquad (14)$$

where $\Phi(u,\sigma_x)$ is the Gaussian function from Eq. (12). Within the barrier, i.e., where $x_0 < x < x_0 + L$, the total evanescent spinor is given by a linear combination of the part of the impinging wave packet that has penetrated the barrier, and is thus *transformed* into an evanescent wave, and the part of the evanescent spinor that is reflected from the right side of the barrier:

$$\varphi_{x_0<x<x_0+L}(x) = \begin{pmatrix} 1 \\ 1 \end{pmatrix} [\alpha e^{-\chi x} + \delta e^{\chi x}], \qquad (15)$$

where $\alpha$ and $\delta$ are numerical coefficients, the squares of which represent probabilities. Finally, beyond the barrier the spinor is represented by a wave packet with reduced amplitude:

$$\psi_{x>x_0+L}(x,t) = c_T \begin{pmatrix} 1 \\ \frac{\gamma}{\gamma+1} \end{pmatrix} \Phi(u,\sigma_x) exp\{-i(kx-\omega t)\}, \qquad (16)$$

where $c_T$ is always a numerical coefficient, whose square represents the probability that the ultrarelativistic particle is transmitted beyond the barrier. Eq. (16) tells us that the particle emerges from the barrier with the same energy that it had prior to tunneling, but does not place constraints to prevent the momentum and rest mass from changing. This peculiarity, although speculative in the context of current knowledge on particle physics, is the point on which the remainder of the theory will be developed.

Because each spinor component must be a continuous, smooth, and differentiable function, the following constraints must hold:

$$\begin{cases} \psi_G(x=x_0,t) = \varphi_G(x=x_0) \\ \psi'_G(x=x_0,t) = \varphi'_G(x=x_0) \\ \psi_G(x=x_0+L,t) = \varphi_G(x=x_0+L) \\ \psi'_G(x=x_0+L,t) = \varphi'_G(x=x_0+L) \end{cases}. \qquad (17)$$

For the purpose of this work, we are interested in calculating the probabilities that the particle penetrates the barrier and that it is transmitted beyond the barrier. This involves calculating the coefficients $\alpha$ and $c_T$. To this end, is necessary to solve the system of linear equations (17), introducing the explicit forms of spinors in (14), (15), and (16) and their derivatives. Concerning the coefficient $\alpha$, the calculation returns

$$\alpha = -2\frac{exp\{im\sqrt{2\gamma_t}c/\hbar\}[\gamma^2(c-u)^2 + \gamma(c-u)\sqrt{2\gamma_t}c]}{[\sqrt{2\gamma_t}c - \gamma(c-u)]^2(e^{\chi L} - e^{-\chi L})}, \tag{18}$$

If $L$ is sufficiently large, then the function $e^{-\chi L}$ goes asymptotically to zero. Under this hypothesis, the probability that the particle is within the barrier in a superluminal state is

$$|\alpha|^2 = 4\frac{[\gamma^2(c-u)^2 + \gamma(c-u)\sqrt{2\gamma_t}c]^2}{[\sqrt{2\gamma_t}c - \gamma(c-u)]^4}. \tag{19}$$

From the conclusions obtained thus far, we know the energy, and therefore the velocity, of the initial particle, but not that of the superluminal state. To obtain the explicit form of the probability (19), we must therefore express the tachyon Lorentz factor $\gamma_t$ as a function of the subluminal one. To this end, we equate Eqs. (1) and (4), representing different explicit forms of the tachyon wave vector. Considering that $E = \gamma mc^2$, we obtain:

$$\gamma_t = \frac{U}{mc^2} - \gamma. \tag{20}$$

Substituting Eq. (20) in Eq. (19) and moving to natural units to simplify the notation, we obtain:

$$|\alpha|^2 = 2\frac{\gamma^2(1-u)^2}{(U-\gamma)}. \tag{21}$$

The probability with which the particle emerges from the barrier is [23]

$$|c_T|^2 = \frac{\gamma^2(1-u)^2(U-\gamma)}{[\gamma^2(1-u)^2 + 2(U-\gamma)]sinh^2\left(L\sqrt{2(U-\gamma)}\right)}. \tag{22}$$

Let us now calculate the ratio $|c_T|^2/|\alpha|^2$, *i.e.*, the ratio between the probability that the particle emerges from the barrier and the probability that it is in a superluminal state:

$$\frac{|c_T|^2}{|\alpha|^2} = \frac{(U-\gamma)^2}{2[\gamma^2(1-u)^2 + 2(U-\gamma)]sinh^2\left(L\sqrt{2(U-\gamma)}\right)}. \tag{23}$$

If $U$ is slightly higher than $E$ and the barrier is sufficiently large, then

$$\lim_{\substack{U \to \gamma \\ L \to \infty}} sinh^2\left(L\sqrt{2(U-\gamma)}\right) \cong \sqrt{2}.$$

Replacing this result in Eq. (23), and considering that $(U - \gamma) \cong 0$, we obtain (returning to conventional units):

$$\frac{|c_T|^2}{|\alpha|^2} = \frac{(U-E)(U/mc^2 - \gamma)}{2\sqrt{2}\gamma^2 m(c-u)^2}. \tag{24}$$

We can view this relationship as the rate of production of an ordinary particle from a tachyon state in the time-lapse during which tunneling occurs. As mentioned above, the tunneling time does not depend on $L$ if the barrier is sufficiently large. We also find this condition in the current theory of neutrino oscillations, based on the hypothesis of flavor mixing [33]. In fact, for an oscillation to occur the length $L$ must be greater than the wavelength of the flavor oscillation [34]. However, if the barrier is sufficiently large, then the tunneling is superluminal. Hence, mass oscillations can occur through a superluminal state. This explains why the study we are developing could help to investigate neutrino flavor oscillations.

Let us return to Eq. (20). Using the explicit forms of the subluminal and superluminal Lorentz factor, we calculate the tachyon velocity associated with the particle state within the barrier:

$$u_t^2 = c^2 \left[\frac{m^2 c^2 (c^2 - u^2)}{U^2}\right]. \tag{25}$$

In the ultrarelativistic regime where $u \cong c$, the tachyon velocity in the barrier is slightly higher than c. Under these conditions, for a barrier with finite dimension and height $U < \infty$ the probability of penetrating the barrier is sufficiently high, but the tachyonic state that is formed is highly energetic. Conversely, if $u \ll c$ then the tachyon velocity is higher than $c$ but the probability of penetration of the barrier is negligible. Finally, by setting an initial value of $u$, we can see that if $U$ is sufficiently small, even if it is greater than $E$, then the term in brackets in Eq. (25) becomes relevant, and the tachyonic velocity assumes a value greater than $c$.

**4. Neutrino Mass Oscillation by Superluminal Tunneling**

Let us now apply the obtained results to neutrinos, with the purpose of explaining flavor oscillations during free flight. Assuming that $U$ is slightly higher than $E$, we can write

$$U = \eta \gamma m c^2 \quad \text{with } \eta > 1,$$

which when substituted into Eq. (24) gives

$$\frac{|c_T|^2}{|\alpha|^2} = \frac{c^2 (\eta - 1)^2}{2\sqrt{2}(c - u)^2}. \tag{26}$$

It has observed that under the assumed approximations, this rate depends exclusively on the velocity of the neutrino arriving at the barrier, and not on its mass. This means that each type of neutrino can undergo a mass oscillation, even a massless one [35-36]. The greater the neutrino velocity and barrier height, the greater the rate (26). It is evident that the barrier height depends on the ability of the neutrino to interact with the matter it traverses. The stronger the interactions are, the greater the height of the equivalent barrier. These interactions [37], mediated by $W$ and $Z^0$ bosons, alter the neutrino kinematics from subluminal to tachyonic. Because of these interactions, the neutrino cannot propagate as it does in free space: everything occurs as if the neutrino crossed a potential barrier. The mediation of the weak interactions, performed by $W$ and $Z^0$ bosons in quantum field theory, is performed by tunneling in relativistic quantum mechanics. However, tunneling also describes superluminal behaviors that are not suggested by the Standard Model, and this makes it possible to study neutrino physics from an unconventional perspective, even if this remains speculative.

Let us now return to the probability that the neutrino is transmitted beyond the potential barrier, as given by Eq. (22). A fraction of this probability describes the neutrino retaining its flavor, while the complement is the probability that the neutrino emerges from the barrier with a different flavor:

$$|c_T|^2 = \rho |c_T|^2 + (1 - \rho)|c_T|^2 \quad 0 \le \rho \le 1. \tag{27}$$

Because the relativistic energy of the transmitted neutrino must be the same as that of the initial one, its change of mass implies a change in momentum. However, the total momentum must be conserved, and this means that the particle changes its momentum during the scattering that takes place inside the barrier. Furthermore, this change must be such that $p^2 c^2 > m^2 c^4$, since the square of the tachyon energy must always be positive. Let us suppose that $\rho |c_T|^2$ is the probability that the neutrino changes flavor, and for simplicity we consider the process $\nu_\mu \to \nu_e$ that occurs when a muon neutrino interacts with matter. The initial state is described by the mixing of the two flavor states. If the matter density is constant, then the probability that the muon neutrino changes flavor is given by [38]

$$P = 1 - \sin^2(2\theta^M) \sin^2\left(\frac{4\pi L}{\lambda_\nu^M}\right). \tag{28}$$

In Eq. (28), $\theta^M$ is the mixing angle in the matter, while $\lambda_\nu^M$ is the oscillation wave length in matter, given by [38]

$$\lambda_\nu^M = \frac{4\hbar cE}{\Delta m^2} \frac{1}{\sqrt{(A/\Delta m^2 - cos2\theta)^2 + sin^2(2\theta)}}. \tag{29}$$

The function $sin^2(2\theta^M)$ is given by [38]

$$sin^2(2\theta^M) = \frac{sin(2\theta)}{\sqrt{(A/\Delta m^2 - cos2\theta)^2 + sin^2(2\theta)}}. \tag{30}$$

Finally, the numerical constant $A$ is

$$A = 2\sqrt{2}(\hbar c)^3 G_F N_e E,$$

where $G_F$ is the Fermi coupling constant, $N_e$ is the electron density in matter, and $E$ is the initial neutrino energy.

Let us now explicitly rewrite the probability of transmission of the neutrino beyond the barrier with a flavor change, using all the approximations adopted in the previous section:

$$\rho|c_T|^2 = \rho \frac{U-E}{mc^2} = \rho(\eta - 1). \tag{31}$$

By equating the probability in (31) with (28), and considering the resonance condition in which the amplitude of oscillation between the two flavor states is maximal (this means that the constant $A$ must be equal to $\Delta m^2 cos2\theta$), we obtain

$$\rho(\eta - 1) = 1 - sin^2\left(4\pi L sin(2\theta)\frac{\Delta m^2}{4\hbar cE}\right). \tag{32}$$

Because $0 \leq \rho \leq 1$, it follows that the constraint on the numerical parameter $\eta$ to achieve resonance is

$$0 \leq \frac{1 - sin^2(4\pi L sin(2\theta)\Delta m^2/4\hbar cE)}{\eta - 1} \leq 1, \tag{33}$$

from which it follows that

$$\eta \geq 2 - sin^2(4\pi L sin(2\theta)\Delta m^2/4\hbar cE). \tag{34}$$

We can then write the barrier energy as a function of the initial neutrino energy and the oscillation parameters:

$$U \geq 2E - E sin^2(4\pi L sin(2\theta)\Delta m^2/4\hbar cE). \tag{35}$$

Because the square sine function oscillates between zero and one, the probability that the superluminal neutrino changes flavor is constrained under the following conditions:

$$\begin{cases} \frac{L}{E} = \frac{K\hbar c}{(2\Delta m^2 sin2\theta)} & \Rightarrow \quad \rho = 0 \\ \frac{L}{E} = \frac{2K\hbar c}{(2\Delta m^2 sin2\theta)} & \Rightarrow \quad \rho \to 1 \end{cases} \quad \text{where } K \text{ is odd}. \tag{36}$$

Eq. (36) provides the initial conditions for having a non-zero probability of neutrino mass oscillation during flight through matter. It is evident that the value of $L$ is essentially determined by the characteristics of the medium crossed by the neutrino, which must be such that its interaction with the surrounding electrons is maximized.

**5. Discussion**

To date, neutrino physics has not found the correct place in the Standard Model: there remain too many open questions to which modern quantum field theory fails to provide satisfactory interpretations, as well described in ref. [38]. Among these, superluminal behavior and flavor oscillations are the most studied and discussed issues. Furthermore, superluminality and flavor oscillation behavior are observed in very similar experiments [39-42], which suggests that they are related to the same physical phenomenon [43-45]. This explains why other authors have attempted to investigate these phenomena by considering them to be intimately connected with each other [43-45].

The purpose of this study is to reconsider these ideas in the framework of relativistic quantum mechanics, taking advantage of the fact that neutrino interactions with matter can be assimilated into the effect of a potential barrier, and that tunneling can lead to superluminal behaviors of particles. As has been proved, within the barrier the neutrino is superluminal, and can decay into one of the possible flavor states. Even if it is based on speculative assumptions, this hypothesis does not violate the laws of quantum mechanics and has already been considered by other influential experts on this subject, even if for the mere study of superluminality [46]. Using a simplified model, in which the initial state is a mix of two flavors, namely muon and electron neutrinos, we have obtained the ratio between the barrier width and particle energy that assures that the resonance condition is satisfied such that the probability of a flavor change is maximized. The problem remains that even if these requirements are satisfied, the neutrino–matter interaction is too weak to be able to obtain an effective potential barrier for superluminal tunneling. This confirms that neutrino superluminality is a local environmental effect [47] and at the same time explains why flavor oscillations are a very rare phenomenon, especially for muon and tau neutrinos. In fact, from Eq. (32) we observe that the probability of oscillation decreases as the term $\Delta m^2$ increases, explaining why the oscillations between tau and muon neutrinos are not favored.

Finally, this theory suggests that research on superluminal neutrinos, for which the few current experimental results have been questioned, must be precisely performed in the context of mass oscillation experiments. The measurement of the arrival time of a muon neutrino from an ultrarelativistic beam of electron neutrinos could effectively contain information on its possible superluminality.

## 6. Conclusion

In this study, it has been shown that a bi-spinor wave packet describing a free spin-1/2 particle in unidirectional motion penetrating a potential barrier is transformed into an evanescent spinor. The components of this spinor are solutions of the tachyon-like KG equation, which proves analytically that the evanescent spinor describes the superluminal behavior of the particle within the barrier. This behavior is also confirmed by the fact that if the barrier is sufficiently large, then the particle tunneling time does not depend on the barrier dimension (the Hartman effect).

The obtained results have been applied to the study of neutrino flavor oscillations. In particular, on the basis of recent experimental results concerning ultrarelativistic neutrinos up to energies of the order of the PeV scale [46], we propose a theory in which flavor changes occur through the *creation* of superluminal states, obtained by the interaction of a neutrino with the surrounding matter. This theory, although still speculative, might be verified by measuring the travel times of only high-energy neutrinos between a source and detector that arrive at the detector with a changed flavor. This theory does not fall within the scope of quantum field theory, but it allows some of the limits that prevent explanations of the anomalous phenomena that characterize neutrino physics, such as flavor oscillations and superluminality, to be remedied.